# AVOIDING BEAM INSTABILITIES AND RESONANCES WITH CIRCULAR MODES*

O. Gilanliogullari[†], Illinois Institute of Technology, Chicago, IL, USA
B. Mustapha, Argonne National Laboratory, Lemont, IL, USA
P. Snopok, Illinois Institute of Technology, Chicago, IL, USA

## Abstract

Beam instabilities and resonances affect the transverse dynamics in particle accelerators and, when encountered, can trigger emittance growth and beam loss. Resonance lines originate from non-linear elements and effects in the lattice, imposing strict constraints on the choice of working points and narrowing the available tune space. Circular modes are round coupled beams with non-zero angular momentum, provide an alternative beam motion and dynamics. In this study, we derive the third-order sextupole resonance conditions in the coupled (normal-mode) parametrization and show that, with circular-mode lattice design and beam operation, most of these resonance lines are naturally suppressed due to the inherent flatness of the mode.

## INTRODUCTION

Accelerators aim to deliver high-quality beams depending on the objectives to be accomplished. The design of an accelerator depends on the beam species to be accelerated and the desired energy and intensity. An accelerator lattice design incorporates both linear and non-linear elements and collective effects. Linear elements are responsible for the total number of betatron oscillations, known as the tunes of the system in orthogonal planes. The inclusion of non-linear elements introduces systematic resonances within the lattice, as discussed in many textbooks [1]. The Courant-Snyder parametrization of the betatron motion [2] shed light on the periodic behavior of the beam and the resonance conditions. Uncontrolled resonances may lead to particle loss; therefore, resonance conditions constrain the selection of tune. Collective effects, such as space charge, also produce resonance conditions [3]. Non-linear static magnetic elements are used for corrections in accelerators. For example, sextupole magnets are used for correcting chromaticity. However, the addition of sextupole elements to the lattice introduces third-order resonances. The resonance diagram shows all possible resonances within a lattice, allowing us to visualize the tune space. In the case of chromatic or collective effects, the tune-shift value is constrained based on the resonance conditions.

Conventional accelerators are designed using the uncoupled Courant-Snyder parametrization [2]. Coupling is typically treated as an error or well-controlled by design. There exist three well-known coupling parametrizations: Edwards-Teng [4], Mais-Ripken [5], and Lebedev-Bogacz [6]. A circular mode beam is a unique type of beam that is inherently flat and has non-zero angular momentum. The dominance of angular momentum in circular modes results in strong coupling within the beam. The intrinsic flatness of the circular mode can be easily converted to real-space flatness, highly desirable for specific applications, by decoupling the beam. However, the early stages of acceleration in rings are affected by space-charge effects in the low-energy region, where flat beams are not preferred. The use of circular modes with round beams at low energy is an attractive solution, enabling flat beams at higher energies through decoupling.

In this paper, we introduce coupling and circular modes expressed using the coupled beam optics formalism. We use both Mais-Ripken and Lebedev-Bogacz parametrizations for coupled beams. We also discuss the resonances arising from the use of high order magnets, such as sextupoles in coupled dynamics and the circular modes case.

## COUPLED BEAM PARAMETRIZATION AND CIRCULAR MODES

The description of a single particle in a coupled system can be expressed in terms of the Lebedev-Bogacz parameterization. This parameterization introduces the eigenvectors of the system and expresses the single-particle coordinates as a combination of projections of eigenmodes. The phase space vector $\vec{z} = [x, x', y, y']^T$ is given by:

$$\vec{z} = \frac{1}{2}\sqrt{2J_1}(\vec{v}_1 e^{-i\psi_1} + \vec{v}_1^* e^{i\psi_1}) + \frac{1}{2}\sqrt{2J_2}(\vec{v}_2 e^{-i\psi_2} + \vec{v}_2^* e^{i\psi_2}), \quad (1)$$

where $J_{1,2}$ are the modes' amplitudes, $\psi_{1,2}$ are the initial eigenmode betatron phases, and $\vec{v}_{1,2}$ are the eigenvectors given as:

$$\vec{v}_1 = \begin{pmatrix} \sqrt{\beta_{1x}} \\ -\frac{i(1-u)+\alpha_{1x}}{\sqrt{\beta_{1x}}} \\ \sqrt{\beta_{1y}} e^{i\nu_1} \\ -\frac{iu+\alpha_{1y}}{\sqrt{\beta_{1y}}} e^{i\nu_1} \end{pmatrix}, \quad \vec{v}_2 = \begin{pmatrix} \sqrt{\beta_{2x}} e^{i\nu_2} \\ -\frac{iu+\alpha_{2x}}{\sqrt{\beta_{2x}}} e^{i\nu_2} \\ \sqrt{\beta_{2y}} \\ -\frac{i(1-u)+\alpha_{2y}}{\sqrt{\beta_{2y}}} \end{pmatrix}, \quad (2)$$

where $\beta_{1x}, \beta_{2x}, \beta_{1y}, \beta_{2y}$ are the four betatron functions, $\alpha_{1x}, \alpha_{2x}, \alpha_{1y}, \alpha_{2y}$ are the alpha or slope functions, $\nu_{1,2}$ are the phases of coupling, and $u$ is the strength of coupling. The phases of coupling determine how eigenmodes project onto the real phase planes, and the coupling strength measures how dominant the coupling is in the lattice. The search for matched solutions for coupled optics is described in detail in [7].

* This work was supported by the U.S. Department of Energy, under Contract No. DE-AC02-06CH11357.
† ogilanli@hawk.iit.edu

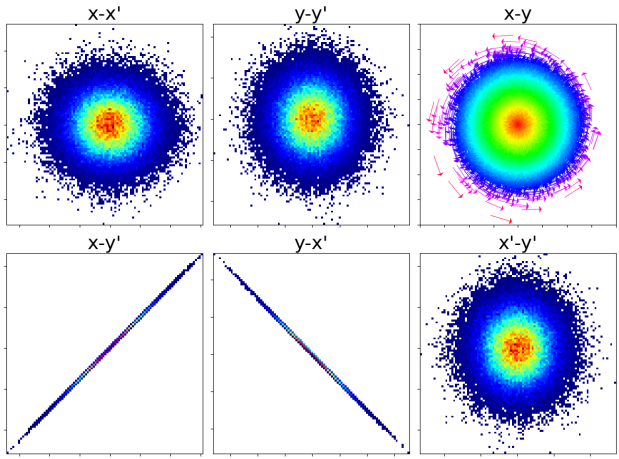

Figure 1: Circular mode beam phase spaces.

The circular mode idea was originally proposed by Y. Derbenev [8] for flat-to-round beam transformers and vortex conditions. The flat-to-round transformer uses three skew quadrupoles to project the real-space flatness onto the $y - x'$ and $x - y'$ phase planes. This transformation and the resulting coupling become clear when one examines the vortex condition. The vortex condition relates the vertical plane $(y, y')$ to the horizontal plane $(x, x')$ as follows:

$$\begin{pmatrix} y \\ y' \end{pmatrix} = \begin{pmatrix} 0 & \beta_c \\ -\frac{1}{\beta_c} & 0 \end{pmatrix} \begin{pmatrix} x \\ x' \end{pmatrix}. \quad (3)$$

Here, $\beta_c$ is the circular mode beta function. The vortex relation suggests that the system is one-dimensional because the $y$ plane is dependent on the $x$ plane. The one-dimensional behavior is projected onto the eigenmode emittances, $\epsilon_2 \ll \epsilon_1$. The dynamics of the system are dominated by eigenmode 1. The phase space coordinates can be written as:

$$x = \sqrt{\epsilon_1 \beta_0} \cos \psi_1, \quad y = \sqrt{\epsilon_1 \beta_0} \sin \psi_1,$$
$$x' = -\frac{1}{2}\sqrt{\frac{\epsilon_1}{\beta_0}} \sin \psi_1, \quad y' = \frac{1}{2}\sqrt{\frac{\epsilon_1}{\beta_0}} \cos \psi_1. \quad (4)$$

The alpha functions are zero, confining the particle to the same path in the transverse plane $x - y$, and using the normalization conditions and projection equations, the phases of couplings $\nu_{1,2} = \pi/2$ and the strength of coupling $u = 1/2$. Circular mode properties also introduce the angular momentum relation to the eigenmode emittances, $L_z = \epsilon_1 - \epsilon_2$. If there is no difference between the eigenmode emittances, then there is no angular momentum. The phase space diagrams for circular modes are shown in Fig. 1. The phase plots of $x - y'$ and $y - x'$ show intrinsic flatness, and the $x - y$ phase space plot shows the transverse momenta of the particles with the direction of motion in the transverse plane. Preservation of circular mode properties, such as angular momentum and roundness, is essential in a design. We have demonstrated in other works [9] that conservation of angular momentum has a direct relation to design phase advances and phases of coupling. The requirement for the conservation of the circular mode is having equal phase advances in both planes. However, this requirement leads to Montague resonance at low energies due to space-charge dominance. To avoid this resonance, solenoids are commonly used for tune splitting and selecting the appropriate tune locations for modes 1 and 2.

## 3RD ORDER RESONANCE

Sextupole magnets are used to correct the unwanted chromatic effects in rings. Due to the non-linearity of the potential, non-linear systematic resonances exist in the ring. The normal sextupole potential is given by

$$V_s = \frac{S}{3}(x^3 - 3xy^2), \quad (5)$$

where $S$ is the sextupole focusing strength. The Lebedev-Bogacz parametrization dictates the coordinates to be expressed as:

$$\begin{aligned} x &= \sqrt{2J_1\beta_{1x}} \cos \psi_1 + \sqrt{2J_2\beta_{2x}} \cos(\psi_2 - \nu_2), \\ y &= \sqrt{2J_1\beta_{1y}} \cos(\psi_1 - \nu_1) + \sqrt{2J_2\beta_{2y}} \cos \psi_2. \end{aligned} \quad (6)$$

Here, $2J_1 = \epsilon_1$ is used. The normal sextupole potential is added to the Hamiltonian written in terms of action-angle variables, and the analysis can be treated as perturbative. Normal-form analysis can also be carried out using Lie algebra methods; for this study, it is not necessary. We assume circular mode optics with phases of coupling $\nu_{1,2} = \pi/2$ that are periodic like the optics functions. Substituting the coordinate parametrization into the sextupole potential yields:

$$\begin{aligned} V_s(\Psi_1, J_1, \Psi_2, J_2) &= A \cos \Psi_1 + B \cos 3\Psi_1 + C \cos(\Psi_1 - 2\Psi_2) \\ &+ D \cos(\Psi_1 + 2\Psi_2) + E \sin(2\Psi_1 - \Psi_2) + F \sin(\Psi_2) \\ &+ G \sin(3\Psi_2) + H \sin(2\Psi_1 + \Psi_2). \end{aligned} \quad (7)$$

Here, $\Psi = \psi - Q\theta$, $\theta = s/R$, and $Q$ is the tune. The amplitudes of the potential are given by

$$\begin{aligned} A &= \frac{S}{\sqrt{2}}(J_1^{3/2}\beta_{1x}^{3/2} - J_1^{3/2}\beta_{1x}^{1/2}\beta_{1y} + 2J_2 J_1^{1/2}\beta_{2x}\beta_{1x}^{1/2} - 2J_2 J_1^{1/2}\beta_{1x}^{1/2}\beta_{2y}), \\ B &= \frac{S}{3\sqrt{2}}(J_1^{3/2}\beta_{1x}^{3/2} + J_1^{3/2}\beta_{1x}^{1/2}\beta_{1y}), \\ C &= -\frac{S}{\sqrt{2}}(J_2 J_1^{1/2}\beta_{1x}^{1/2}\beta_{2x} + J_2 J_1^{1/2}\beta_{1x}^{1/2}\beta_{2y} - 2J_1^{1/2}J_2\beta_{1y}^{1/2}\beta_{2x}^{1/2}\beta_{2y}^{1/2}), \\ D &= -\frac{S}{\sqrt{2}}(J_2 J_1^{1/2}\beta_{1x}^{1/2}\beta_{2x} + J_2 J_1^{1/2}\beta_{1x}^{1/2}\beta_{2y} - 2J_1^{1/2}J_2\beta_{1y}^{1/2}\beta_{2x}^{1/2}\beta_{2y}^{1/2}), \\ E &= -\frac{S}{\sqrt{2}}(J_1 J_2^{1/2}\beta_{1x}\beta_{2x}^{1/2} + J_1 J_2^{1/2}\beta_{1y}\beta_{2x}^{1/2} + 2J_1 J_2^{1/2}\beta_{1x}^{1/2}\beta_{1y}^{1/2}\beta_{2y}^{1/2}), \\ F &= \frac{S}{\sqrt{2}}(2J_1 J_2^{1/2}\beta_{1x}\beta_{2x}^{1/2} - 2J_1 J_2^{1/2}\beta_{1y}\beta_{2x}^{1/2} + J_2^{3/2}\beta_{2x}^{3/2} - J_2^{3/2}\beta_{2x}^{1/2}\beta_{2y}), \\ G &= -\frac{S}{3\sqrt{2}}(J_2^{3/2}\beta_{2x}^{3/2} + 3J_2^{3/2}\beta_{2y}\beta_{2x}^{1/2}), \\ H &= \frac{S}{\sqrt{2}}(J_1 J_2^{1/2}\beta_{1x}\beta_{2x}^{1/2} + J_1 J_2^{1/2}\beta_{1y}\beta_{2x}^{1/2} - 2J_1^{1/2}J_2^{1/2}\beta_{1x}^{1/2}\beta_{1y}^{1/2}\beta_{2y}^{1/2}). \end{aligned} \quad (8)$$

Table 1 shows the resonances from the sextupole perturbation. Interestingly, there are eight resonance conditions from coupled parametrization, whereas uncoupled dynamics have only four. Setting off-mode functions, $\beta_{2x} = \beta_{1y} = 0$, recovers the uncoupled dynamics resonances. The comparison of the resonance diagrams between the coupled and uncoupled cases is shown in Fig. 2.

Table 1: Resonances Due to Sextupoles ($\nu_{1,2} = \pi/2$)

| Resonance | Driving Term | Amplitude |
| --- | --- | --- |
| $Q_1 = n$ | $\cos(\Psi_1)$ | $A(J_1, J_2)$ |
| $3Q_1 = n$ | $\cos(3\Psi_1)$ | $B(J_1, J_2)$ |
| $Q_2 = n$ | $\sin(\Psi_2)$ | $F(J_1, J_2)$ |
| $3Q_2 = n$ | $\sin(3\Psi_2)$ | $G(J_1, J_2)$ |
| $Q_1 - 2Q_2 = n$ | $\cos(\Psi_1 - 2\Psi_2)$ | $C(J_1, J_2)$ |
| $Q_1 + 2Q_2 = n$ | $\cos(\Psi_1 + 2\Psi_2)$ | $D(J_1, J_2)$ |
| $2Q_1 - Q_2 = n$ | $\sin(2\Psi_1 - \Psi_2)$ | $E(J_1, J_2)$ |
| $2Q_1 + Q_2 = n$ | $\sin(2\Psi_1 + \Psi_2)$ | $H(J_1, J_2)$ |

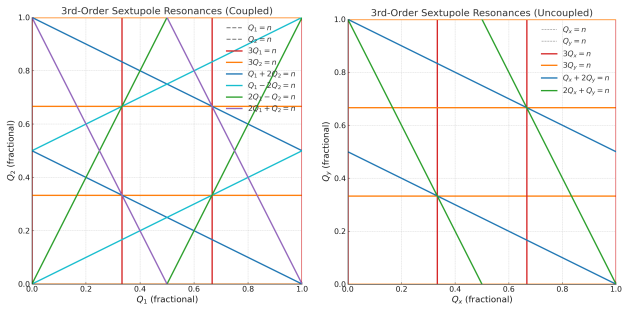

Figure 2: Tune diagrams showing sextupole resonances: Coupled parametrization (left), Uncoupled parametrization (right).

The amplitudes of the sextupole potential depend on both actions, $J_1$ and $J_2$. The main property of circular modes is intrinsic flatness, $J_2 \ll J_1$. Ideally, mode 2 eigenmode emittance is much smaller than that of mode 1. This suggests that, for circular modes, mode 2 does not contribute to the dynamics. Setting $J_2 = 0$ removes many of the amplitudes and yields the terms $A$ and $B$, which are associated with integer and third integer resonances.

Using the formalism described in [7], we designed a small ring that puts one of the tunes on resonance, as shown in Fig. 3. The tune splitting is controlled by the solenoids in the straight sections. Detailed beam simulations were performed for a 0.3 MeV/u gold ions with $\epsilon_1 = 1.0$ mm-mrad and $\epsilon_2 = 0.1$ µm-mrad beam. The tunes are controlled using solenoids, and $Q_1 = 3.945$ and $Q_2 = 3.333$. Reversing solenoid polarity puts the dominant mode 1 on resonance. Switching solenoid polarity allows us to switch between two resonance modes. The gold ions are tracked in this storage ring using ImpacT-X tracking simulation [10]. Tracking particles through a round circular aperture with a 3-cm radius for beam stability effects.

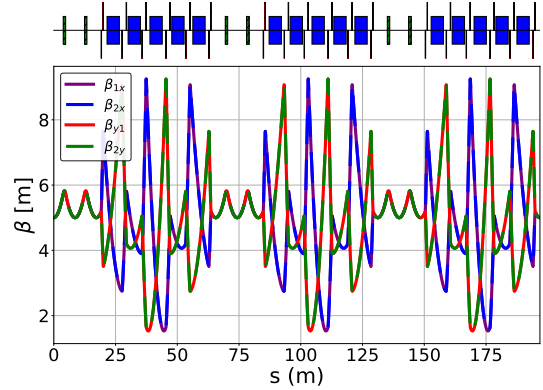

Figure 3: Lattice design with mode 2 on tune.

The operating points of the lattice are shown in Fig. 4. The red star corresponds to mode 1 being on resonance, while the green cross corresponds to mode 2 being on resonance. The number of particles that survived through ten thousand turns is shown in Fig. 5. Surviving particles indicate that even if mode 2 is on resonance, the beam does not experience the resonance due to its intrinsic flatness, which is an interesting phenomenon and confirms that circular modes behave like one-dimensional beams.

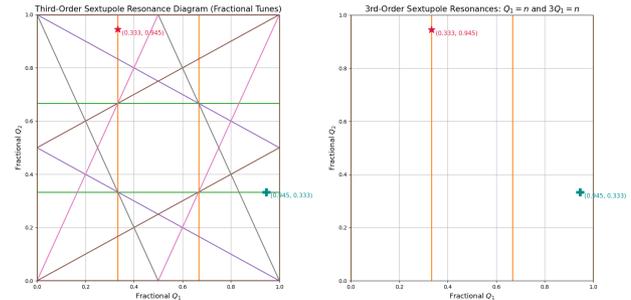

Figure 4: Working points of the lattice.

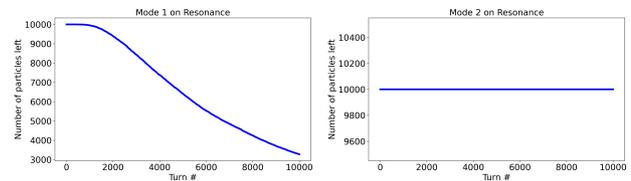

Figure 5: Number of particles survived within aperture of 3 cm. Mode 1 on resonance (left). Mode 2 on resonance (right).

## CONCLUSION

In conclusion, we have shown that in coupled dynamics for third-order resonance, there exist twice as many resonance conditions as in the uncoupled case. Circular mode beams reduce the eight resonance conditions to two. We also show that putting a less dominant mode on resonance does not trigger particle loss as it does for the dominant mode.